\documentclass[a4paper,12pt]{article}
\usepackage[utf8x]{inputenc}
\usepackage{amsmath,amssymb,cite}
\usepackage{ dsfont }
\usepackage{tikz}
\usepackage{graphics}

\title{Time dependent transitions with time-space noncommutativity $\&$ its implications in Quantum Optics}
\author{Nitin Chandra\footnote{nitin@cts.iisc.ernet.in} \\
\begin{small}
Centre for High Energy Physics, Indian Institute of Science, Bangalore, India.
\end{small}}
\date{\empty}
\begin{document}

\maketitle

\begin{abstract}
We study the time dependent transitions of quantum forced harmonic oscillator (QFHO) in noncommutative $\mathds{R}^{1,1}$
perturbatively to linear order in the noncommutativity $\theta$.
We show that the Poisson distribution gets modified, and that the vacuum state evolves into a ``squeezed'' state 
rather than a coherent state.
The time evolutions of uncertainties in position and momentum in vacuum are also studied 
and imply interesting consequences for modeling nonlinear phenomena in quantum optics.
\end{abstract}
\section{Introduction}
There seems to be a growing consensus among physicists that our classical notion of spacetime has to be drastically revised 
in order to find a consistent formulation of quantum mechanics and gravity \cite{Doplicher:1994tu, Seiberg:2006wf, Rovelli:2003cu}.
One possible generalization that has attracted much interest is that of noncommutative Moyal spacetime 
\cite{Douglas:2001ba, Seiberg:1999vs, Chaichian:2004yh,Balachandran:2008gr, Balachandran:2006pi, Lizzi:2006xi, Balachandran:2007kv, Basu:2010qm, Acharyya:2010yf, Szabo:2001kg, Balachandran:2010wc}.
In situations where the time coordinate remains commutative, i.e., only the spatial coordinates do not commute with each other,
the quantum theory is conceptually straightforward (but nonetheless may display novel phenomena)
\cite{Dragovich:2003sj, Scholtz:2008zu, Ho:2001aa, Bolonek:2002cc, BenGeloun:2009zd, Rohwer:2010zq, Scholtz:2007ig}.
In this article we will concentrate on understanding some implications of quantum mechanics with time-space noncommutativity,
specifically we will work with the Moyal plane $\mathds{R}^{1,1}_\theta$.
We will use the formalism of unitary quantum mechanics on this space as developed by Balachandran et. al. \cite{Balachandran:2004rq} 
(see also \cite{Li:2001vc}).

When time and space do not commute with each other it is not unreasonable to expect that the dynamics of the time dependent 
processes get altered.
We will verify this explicitly in the context of a simple model of the forced harmonic oscillator (FHO) with the forcing term 
switched on only for a finite duration of time.
In the commutative case this is a much studied model.
We will compute deviations from the commutative case to leading order in $\theta$.
These deviations suggest that time-space noncommutativity can capture certain nonlinear effects seen in quantum optics.
This article is organized as follows: In section \ref{bala_paper} we will briefly review the formulation of unitary 
quantum mechanics on $\mathds{R}^{1,1}_\theta$ \cite{Balachandran:2004rq}. 
In section \ref{problem} we will solve the problem of the FHO perturbatively in $\theta$ and compute corrections to the 
transition probabilities between simple harmonic oscillator (SHO) states.
These corrections suggest the noncoherent nature of the time-evolved vacuum state 
and are the reminiscent of those seen in nonlinear quantum optics \cite{Walls_book}.
To flesh out this analogy better we study the time-evolution of uncertainties 
in position and momentum in section \ref{uncertainties}.
Encouraged by these results we, in section \ref{quantum_optics}, suggest a correspondence between the nonlinearity in 
quantum optics and the quantum mechanics on $\mathds{R}^{1,1}_\theta$.
We conclude with a summary of our results in section \ref{conclusions}.

\section{Unitary Quantum Mechanics on $\mathds{R}^{1,1}_\theta$}\label{bala_paper}
The noncommutative space $\mathds{R}^{1,1}_\theta$ is described by the coordinates $\hat{x}_\mu$'s satisfying
\begin{equation} \label{commutator}
 [\hat{x}_\mu,\hat{x}_\nu]=i\theta\varepsilon_{\mu\nu}\mbox{ with } 
\varepsilon_{\mu\nu}=-\varepsilon_{\nu\mu}\mbox{ and } \varepsilon_{01}=1,
\end{equation}
where $\mu$ and $\nu$ can take values 0,1.
Without loss of generality we can take $\theta>0$, as its sign can always be flipped by changing $\hat{x}_1$ to $-\hat{x}_1$.  
Let $\mathcal{A}_{\theta}(\mathds{R}^{1,1})$ be the unital algebra generated by $\hat{x}_0$ and $\hat{x}_1$.
We associate to each  $\hat{\alpha}\in\mathcal{A}_{\theta}\left(\mathds{R}^{1,1}\right)$,
 its left and right representations $\hat{\alpha}^{L}$ and $\hat{\alpha}^{R}$:
\begin{equation}
\hat{\alpha}^{L}\hat{\beta} = \hat{\alpha}\hat{\beta} \,\, , \,\,
\hat{\alpha}^{R}\hat{\beta} = \hat{\beta}\hat{\alpha} \,\, , \,\,
\hat{\beta}\in\mathcal{A}_{\theta}\left(\mathds{R}^{1,1}\right) \,\,.
\label{Left_Right_reps}
\end{equation}
Unless stated, we work with the left representation. \\
\indent For a quantum theory, what we need are:
(1) a suitable inner product on 
$\mathcal{A}_{\theta}\left(\mathds{R}^{1,1}\right)$;
(2) a Schr\"{o}dinger constraint on
$\mathcal{A}_{\theta}\left(\mathds{R}^{1,1}\right)$;
and (3) a self-adjoint (with respect to the inner product defined) Hamiltonian $\hat{H}$ and observables which act on the
constrained subspace of $\mathcal{A}_{\theta}\left(\mathds{R}^{1,1}\right)$. \\
{\it 1. The Inner Product:} \\
There are several suitable inner products and they are all equivalent to each other.
One example is given here:
We associate a symbol $\alpha_S$ corresponding to each $\hat{\alpha} = \int d^{2}k \,
\tilde{\alpha}(k) e^{ik_{1}\hat{x}_{1}}e^{ik_{0}\hat{x}_{0}} \in\mathcal{A}_\theta(\mathds{R}^{1,1})$ as
\begin{equation}
\alpha_{S}(x_{0},x_{1}) = \int d^{2}k \,
\tilde{\alpha}(k) e^{ik_{1}x_{1}}e^{ik_{0}x_{0}} \,\, .
\label{alpha-s-symbol}
\end{equation}
Note that $x_0$ and $x_1$ and hence $\alpha_S$ are purely commutative.
The inner product is defined as
\begin{equation}
\left(\hat{\alpha},\hat{\beta}\right)_{t} =
\int dx_{1} \, \alpha_{S}^{*}(t,x_{1})\beta_{S}(t,x_{1}) \,\, .
\label{S-inner-product}
\end{equation}
{\it 2. The Schr\"{o}dinger Constraint and time evolution:} \\
The operators $\hat{P}_0$ and $\hat{P}_1$, given by 
\begin{equation}
i\frac{\partial}{\partial x_{0}}\equiv\hat{P}_{0} =
-\frac{1}{\theta}{\rm ad}\,\hat{x}_{1}, \quad
-i\frac{\partial}{\partial x_{1}}\equiv\hat{P}_{1} =
-\frac{1}{\theta}{\rm ad}\,\hat{x}_{0},
\end{equation}
generate time and space translations respectively.
The Hamiltonian $\hat{H}$, in general, may depend on $\hat{x}_1^L, \hat{x}_0^R$ and $\hat{P}_1$.
The possible dependence of $\hat{x}_1^R$ and $\hat{x}_0^L$ can be bypassed by
\begin{equation}
 \hat{x}_{1}^{R} = \theta\hat{P}_0+\hat{x}_{1}^{L}, \quad
\hat{x}_{0}^{L} = \theta \hat{P}_{1} + \hat{x}_{0}^{R}. \label{lr0_relation}
\end{equation}
Also, there is no dependence on $\hat{P}_0$ assumed in the line of the commutative case 
where there is never such dependence of H on $i\partial_{x_0}$ for $\theta = 0$.
Now note that the inner product (\ref{S-inner-product}) has an explicit dependence on the parameter $t$ and hence there exist more than one 
null vectors with respect to this inner product (actually any vector which vanishes at $\hat{x}_0 = t$ is a null vector).
But this fact need not bother us as we are only interested in those states that satisfy the Schr\"{o}dinger constraint
\begin{equation}
\left(\hat{P}_0-\hat{H} \right) \hat{\psi} = 0.
\end{equation}
It is easy to see that now there are no non-trivial null vectors.
The Hamiltonian $\hat{H}$ depends on $\hat{x}_1^L, \hat{x}_0^R$ and $\hat{P}_1$.
Since $\hat{x}_0^R$ commutes with $\hat{x}_1^L$ and $\hat{P}_1$ we will choose $\hat{x}_0^R$ as ``time''.

It is easy to write down the formal solution of the Schr\"{o}dinger constraint and find the time evolution.
The time evolution is given by $\hat{x}_0\rightarrow \hat{x}_0+\tau$ (or equivalently by $\hat{x}_0^R\rightarrow \hat{x}_0^R+\tau$).
Thus the amount of time-translation is always commutative, though the time-operator itself is noncommutative.
The time evolved wavefunctions satisfying the Schr\"{o}dinger constraint are of the form
$\hat{\psi}(\hat{x}_0,\hat{x}_1)=\hat{U}\left(\hat{x}_{0}^{R},\tau_{I}\right)\hat{\chi}(\hat{x}_1)$, where
\begin{equation}
\hat{U}\left(\hat{x}_{0}^{R},\tau_{I}\right) =
\left(\left. T \exp\left[-i\left(\int_{\tau_{I}}^{x_{0}}
d\tau\,\hat{H}\left(\tau,\hat{x}_{1}^{L},\hat{P}_{1}
\right) \right) \right]\right) \right|_{x_{0} =
\hat{x}_{0}^{R}} \,\,.
\label{formula_psi_2}
\end{equation}
{\it 3. The Spectral Map:} \\
Consider a time-independent Hamiltonian $\hat{H} = \frac{\hat{P}_{1}^{2}}{2m}+V(\hat{x}_{1})$.
The corresponding commutative Hamiltonian is $H = -\frac{1}{2m}\frac{\partial^{2}}{\partial x_{1}^{2}}+V(x_{1})$,
with eigenfunctions $\psi_{E}\left(x_{0},x_{1}\right) = \varphi_{E}(x_{1})e^{-iE x_{0}}$ and eigenvalues E. 
The spectrum of the corresponding noncommutative $\hat{H}$ will be given by
$\hat{\psi}_{E}=e^{-iE\hat{x}_{0}^R}\varphi_{E}(\hat{x}_{1})=\varphi_{E}(\hat{x}_{1})e^{-iE\hat{x}_{0}}$
with the same eigenvalues $E$ as
$\hat{H}\varphi_{E}(\hat{x}_{1})=E\varphi_{E}(\hat{x}_{1})$.
Here $\varphi_E(\hat{x}_1)$ has been obtained by replacing $x_1$ with $\hat{x}_1$ in $\varphi_E(x_1)$.

\section{QFHO in $\mathds{R}_\theta^{1,1}$  and their Transition Probabilities} \label{problem}
Let us recall the dynamics of a QFHO in ordinary spacetime. The Hamiltonian of this system is given by
\begin{equation}
H(t)=\frac{p^2}{2m_0}+\frac{1}{2}m_0\omega^2 x^2+f(t)x+g(t)p,
\label{cfho}
\end{equation} 
where $m_0$ is the mass of the particle and $\omega$ is the angular frequency of the oscillator.
We are interested in real functions obeying 
\begin{equation}
f(t),g(t)=0\mbox{   for } t\rightarrow\pm\infty.
\label{bc}
\end{equation}
At $t\rightarrow -\infty$ the Hamiltonian is simple harmonic 
and we assume the system to be in one of the eigenstates of this SHO Hamiltonian.
At $t\rightarrow \infty$ the Hamiltonian again becomes simple harmonic
and we try to find the probability (the transition probability) for the system to be in any arbitrary eigenstate of 
the SHO Hamiltonian subjected to the fact that the system was in some already given eigenstate at $t\rightarrow -\infty$.
For this what we do is the following:
\begin{itemize}
\item First we assume our system to be in an eigenstate $\phi_n(x)$ at $t=t_i\rightarrow -\infty$.
\item The state $\phi_n(x)$ evolves under the SHO Hamiltonian from $t=t_i\rightarrow -\infty$ to $t=T_1$.
\item At $t=T_1$ the interaction gets switched on.
\item The system then evolves under the full Hamiltonian (\ref{cfho}) from $t=T_1$ to $t=T_2$.
\item At $t=T_2$ the interaction gets switched off.
\item The system again evolves under the SHO Hamiltonian from $t=T_2$ to $t=t_f\rightarrow \infty$.
\item We find the inner product of the final state we get at $t=t_f\rightarrow \infty$ with the eigenstate $\phi_m(x)$.
This gives the Transition Amplitude $A_{mn}$ while its absolute square gives the Transition Probability $P_{mn}$.
\end{itemize}
The generalization of the above Hamiltonian in $\mathds{R}_\theta^{1,1}$ is
\begin{equation}
\hat{H} = \frac{\hat{p}_1^2}{2m_0}+\frac{1}{2}m_0\omega^2\hat{x}_1^2
+\frac{1}{2}[f(\hat{x}_0)\hat{x}_1+\hat{x}_1f(\hat{x}_0)]+g(\hat{x}_0)\hat{p}_1
 = \hat{H}_0+\hat{H}_I,
\label{fho}
\end{equation}
with
\begin{equation}
\hat{H}_0 = \frac{\hat{p}_1^2}{2m_0}+\frac{1}{2}m_0\omega^2\hat{x}_1^2, \quad
\hat{H}_I = \frac{1}{2}[f(\hat{x}_0)\hat{x}_1+\hat{x}_1f(\hat{x}_0)]+g(\hat{x}_0)\hat{p}_1.
\label{h_0_i}
\end{equation}
As $\hat{x}_0$ and $\hat{p}_1$ commute with each other, the ordering does not matter in the last term.\\
\indent To define the transitions for the above Hamiltonian consider the time evolution by an amount $\tau$.
The functions $f(\hat{x}_0)$ and $g(\hat{x}_0)$ have the properties of vanishing in the far past and the far future, i.e.,
\begin{equation}
f,g(\hat{x}_0+\tau)\rightarrow 0 \mbox{\,\,\, \rm{as}\,\,\,} \tau\rightarrow \pm \infty.
\end{equation}
We shall find the transition probabilities ($P_{m,n}$) for an SHO state ``$n$'' 
at initial time ($\tau\rightarrow -\infty$) to go to some other SHO state ``$m$'' at final time ($\tau\rightarrow +\infty$) 
after evolving under the Hamiltonian (\ref{fho}).
The Spectral Map tells us that the energy spectrum of the SHO Hamiltonian in $\mathds{R}_\theta^{1,1}$ is same as 
that of the commutative one, i.e., 
\begin{equation}
E_n = \hbar\omega\left(n+\frac{1}{2}\right), \quad  
\psi_n(\hat{x}_0,\hat{x}_1) = \phi_n(\hat{x}_1)e^{-i\omega\left(n+\frac{1}{2}\right)\hat{x}_0},
\label{En_psin}
\end{equation}
where $\phi_n(x_1)$ is the eigenfunctions of the commutative SHO Hamiltonian.
The orthonarmality (apart from a phase factor which comes because of the time evolution) of the eigenfunctions 
$\psi_n(\hat{x}_0,\hat{x}_1)$ with respect to the inner product defined in section \ref{bala_paper} can easily be checked.
The transition probabilities for our problem can be found by computing the same 
for the commutative Hamiltonian obtained after replacing
\begin{equation}
\hat{x}_0  \rightarrow -\frac{\theta}{\hbar} p+t,\quad \hat{x}_1 \rightarrow  x , \quad \hat{p}_1  \rightarrow  p
\label{replace}
\end{equation}
in the Hamiltonian (\ref{fho}).
Here $t$ has come in place of the ``time'' $\hat{x}_0^R$ which commutes with $\hat{x}_1$ and $\hat{p}_1$.
To linear order in $\theta$ we obtain the following commutative Hamiltonian
\begin{equation}
H(t)= H_0 + H_I(t) = H_0 + H_{I0}(t)+\theta H_{I1}(t),
\label{h}
\end{equation}
with
\begin{equation} \left.
\begin{array}{l}
H_0 = \frac{p^2}{2m_0}+\frac{1}{2}m_0\omega^2 x^2 
 = \hbar\omega\left(a^\dagger a+\frac{1}{2}\right),\quad  
H_{I0} = f(t)x+g(t)p = z^*(t)a+z(t)a^\dagger \\
H_{I1} = \frac{1}{\hbar}\left(-g^{\prime}(t)p^2-\frac{1}{2}f^{\prime}(t)(xp+px)\right) = 
\frac{i}{\hbar}\sqrt{\frac{m_0\hbar\omega}{2}}\left(z^{*\prime}(t)a^2-z^{\prime}(t)a^{\dagger 2}+
i\sqrt{\frac{m_0\hbar\omega}{2}}g^{\prime}(t)(2a^{\dagger} a+1)\right)
\end{array}\right\}.
 \label{h0}
\end{equation}
The function $z(t)$ is related to $f(t)$ and $g(t)$ as
\begin{equation}
z(t)=\sqrt{\frac{\hbar}{2m_0\omega}}\left(f(t)+im_0\omega g(t)\right).
\label{z}
\end{equation}
Also, $a$ and $a^\dagger$ are the annihilation and creation operators respectively defined as
\begin{equation*}
\begin{array}{rcl}
a=\sqrt{\frac{m_0\omega}{2\hbar}}\left(x+i\frac{p}{m_0\omega}\right) 
&& x=\sqrt{\frac{\hbar}{2m_0\omega}}\left(a^\dagger+a\right) \\
&\Rightarrow& \\
a^\dagger=\sqrt{\frac{m_0\omega}{2\hbar}}\left(x-i\frac{p}{m_0\omega}\right) 
&& p=i\sqrt{\frac{m_0\hbar\omega}{2}}\left(a^\dagger-a\right)
\end{array}
\end{equation*}
The nonlinearity in the Hamiltonian (\ref{h}) is purely due to the noncommutativity.
This provokes us to model certain types of nonlinear phenomena in quantum optics by the noncommutativity between time and space coordinates.
This analogy will be further studied in section \ref{quantum_optics}. 
Let us now continue with calculating the transition amplitude which is given by
\begin{equation}
A_{m,n}(t_f,T_2;T_1,t_i)=\langle \phi_m|U_0^{\dagger}(t_f,t_i)U_0(t_f,T_2)U(T_2,T_1)U_0(T_1,t_i)|\phi_n\rangle,
\label{A}
\end{equation}  
where $U_0(t^{\prime},t)$ and $U(t^{\prime},t)$ are the time evolution operators from time $t$ to time $t^{\prime}$ for 
the Hamiltonians $H_0$ and $H(t)$ respectively, i.e.,
\begin{equation}
 U_0(t^{\prime},t) = e^{-\frac{i}{\hbar}H_0(t^{\prime}-t)}, \quad  
U(t^{\prime},t) = T\left[e^{-\frac{i}{\hbar}\int_{t}^{t^{\prime}}
d\tau H(\tau)}\right],
\end{equation}
the latter one being the time-ordered exponential.
This gives
\begin{equation}
A_{m,n}(t_f,T_2;T_1,t_i)=e^{\frac{i}{\hbar}[E_m(T_2-t_i)+E_n(t_i-T_1)]}\langle\phi_m|U(T_2,T_1)|\phi_n\rangle.
\label{A_}
\end{equation}
The state $|\psi(t)\rangle = U(t,T_1)|\phi_n\rangle$ evolves according to the Schr\"{o}dinger equation for the Hamiltonian (\ref{h}) 
\begin{equation}
\left( i\hbar \frac{d}{dt}-H_0\right)|\psi(t)\rangle = H_I(t)|\psi(t)\rangle,
\label{Schro}
\end{equation}
with the initial condition $|\psi(t=T_1)\rangle=|\phi_n\rangle$.
If we define the Green's operator function $G(t,t_0)$ as
\begin{equation}
\left( i\hbar \frac{\partial}{\partial t}-H_0\right)G(t,t_0)=\delta (t,t_0),
\label{G}
\end{equation}
then solution of the Schr\"{o}dinger  equation (\ref{Schro}) will be
\begin{equation*}
|\psi(t)\rangle =|\phi(t)\rangle +\int_{-\infty}^{+\infty} dt_0 \,\,G(t,t_0)H_I(t_0)|\psi(t_0)\rangle,
\end{equation*}
which in turn gives the Born series
\begin{eqnarray}
|\psi(t)\rangle &=& |\phi(t)\rangle +\int_{-\infty}^{+\infty} dt_0 G(t,t_0)H_I(t_0)|\phi(t_0)\rangle \nonumber \\
&&+\int_{-\infty}^{+\infty} dt_0 \int_{-\infty}^{+\infty} dt_1 G(t,t_0)H_I(t_0)G(t_0,t_1)H_I(t_1)|\phi(t_1)\rangle+.... \label{Born}
\end{eqnarray}
Here $|\phi(t)\rangle$ is the solution of the homogeneous equation $\left( i\hbar \frac{d}{dt}-H_0\right)|\phi(t)\rangle=0$,
which is nothing but the Schr\"{o}dinger  equation for SHO.
$G$ has been found in the Appendix \ref{greens} (see (\ref{G_sol})).
Note that the $\Theta$-function in the expression of the $G$  restricts the integration over $t_j$ in (\ref{Born}) 
within the limit of $-\infty$ to $t_{j-1}$ ($t_{-1}=t$).
Thus, at $t=T_1$ the integrations are only in the intervals when the interaction was switched off, i.e., $H_I=0$.
Hence, we get $|\psi(t=T_1)\rangle=|\phi(t=T_1)\rangle=|\phi_n\rangle$.
The solution of the homogeneous part $|\phi(t)\rangle$ with this initial condition is 
\begin{equation}
|\phi(t)\rangle=e^{-\frac{i}{\hbar}E_n(t-T_1)}|\phi_n\rangle.
\end{equation}
Now, putting (\ref{Born}) with $t=T_2$ for $U(T_2,T_1)|\phi_n\rangle$ in (\ref{A_}), we get
\begin{equation} \label{A__}
A_{m,n}(t_f,T_2;T_1,t_i)=\sum_{j=0}^{\infty}B_j(t_f,T_2;T_1,t_i),
\end{equation}
with
\begin{equation}\left.
\begin{array}{rcl}
B_0(t_f,T_2;T_1,t_i) &=& \delta_{m,n} \\
B_j(t_f,T_2;T_1,t_i) &=& \int_{-\infty}^{+\infty}dt_0\int_{-\infty}^{+\infty}dt_1...\int_{-\infty}^{+\infty}dt_{j-1} F_{m,n}^{j}(t_f,T_2;t_0,t_1,...,t_{j-1};T_1,t_i) \\
\end{array}\right\rbrace
\end{equation}
for $j=1,2,...$.
Here
\begin{equation}
\begin{array}{rcl}
 F_{m,n}^{j}(t_f,T_2;t_0,t_1,...,t_{j-1};T_1,t_i) &=& 
\left(-\frac{i}{\hbar}\right)^j\Theta (T_2-t_0)\Theta (t_0-t_1)...\Theta (t_{j-2}-t_{j-1}) \\
&& e^{\frac{i}{\hbar}(E_n-E_m)t_i}\langle \phi_m|H_I^{int}(t_0)H_I^{int}(t_1)...H_I^{int}(t_{j-1})|\phi_n\rangle.
\end{array}
\end{equation}
The $H^{int}$'s are defined as
\begin{equation}
H_{(...)}^{int}(t) = e^{\frac{i}{\hbar}H_0t}H_{(...)}(t)e^{-\frac{i}{\hbar}H_0t}
\end{equation}
Separating the $\theta$-dependent and independent part we get
\begin{equation}
A_{m,n}(t_f,T_2;T_1,t_i)=e^{\frac{i}{\hbar}(E_n-E_m)t_i}\langle \phi_m|[A^{(0)}(T_2,T_1)+\theta A^{(1)}(T_2,T_1)]|\phi_n\rangle,
\label{Amn}
\end{equation}
with
\begin{equation}
\begin{array}{rcl}
A^{(0)}(T_2,T_1) &=& \mathbf{I}+\int_{-\infty}^{+\infty}dt_0\left( -\frac{i}{\hbar}\right) \Theta (T_2-t_0)H_{I0}^{int}(t_0)\\
&&+\int_{-\infty}^{+\infty}dt_0\int_{-\infty}^{+\infty}dt_1\left( -\frac{i}{\hbar}\right)^2 \Theta (T_2-t_0)\Theta (t_0-t_1)H_{I0}^{int}(t_0)H_{I0}^{int}(t_1) \\
&& +... \\
&=& T\left[ e^{-\frac{i}{\hbar}\int_{-\infty}^{\infty}d\tau H_{I0}^{int}(\tau)}\right] \label{A0},
\end{array}
\end{equation}
\begin{equation} \label{A1_series}
\begin{array}{rl}
A^{(1)}(T_2,T_1)]=& -\frac{i}{\hbar}\int_{-\infty}^{+\infty}dt_0\Theta (T_2-t_0)H_{I1}^{int}(t_0) \\
\\
& +\left(-\frac{i}{\hbar}\right)^2\int_{-\infty}^{+\infty}dt_0\int_{-\infty}^{+\infty}dt_1\Theta (T_2-t_0)\Theta (t_0-t_1)
.[H_{I1}^{int}(t_0)H_{I0}^{int}(t_1) + H_{I0}^{int}(t_0) H_{I1}^{int}(t_1)] \\
\\
& +\left(-\frac{i}{\hbar}\right)^3\int_{-\infty}^{+\infty}dt_0\int_{-\infty}^{+\infty}dt_1\int_{-\infty}^{+\infty}
dt_2\Theta (T_2-t_0)\Theta (t_0-t_1)\Theta (t_1-t_2) \\
\\
& \,\,\,\,\, .[H_{I1}^{int}(t_0)H_{I0}^{int}(t_1)H_{I0}^{int}(t_2) + H_{I0}^{int}(t_0)H_{I1}^{int}(t_1)H_{I0}^{int}(t_2)
+ H_{I0}^{int}(t_0)H_{I0}^{int}(t_1)H_{I1}^{int}(t_2)] \\
\\
& +....
\end{array}
\end{equation}
The above expression for $A^{(1)}(T_2,T_1)$ can be simplified to (see Appendix \ref{simplify_A1}) 
\begin{equation} \label{A1_A0}
A^{(1)}(T_2,T_1)=-\frac{i}{\hbar}A^{(0)}(T_2,T_1)\int_{-\infty}^{\infty}dt_0
\left[A^{(0)}(t_0,T_1)\right]^{-1}H_{I1}^{int}(t_0)A^{(0)}(t_0,T_1).
\end{equation}
$A^{(0)}(t,t^{\prime})$ with arbitrary arguments is defined in (\ref{Gint_sol}).
Putting this in equation (\ref{Amn}) we get
\begin{equation}
A_{m,n}(t_f,T_2;T_1,t_i)=
e^{\frac{i}{\hbar}(E_n-E_m)t_i}\langle \phi_m|A^{(0)}\left[\mathbf{I}-\frac{i}{\hbar}\theta 
\int_{-\infty}^{\infty}dt_0[A^{(0)}(t_0,T_1)]^{-1}H_{I1}^{int}(t_0)A^{(0)}(t_0,T_1)\right]|\phi_n\rangle.
\label{Amn_A0}
\end{equation}
A straightforward use of the identity
\begin{equation}
e^{\lambda A}Be^{-\lambda A}=B+\frac{\lambda}{1!}[A,B]+\frac{\lambda^2}{2!}\left[A,[A,B]\right]+\frac{\lambda^3}{3!}\left[A,\left[A,[A,B]\right]\right]+...
\label{identity}
\end{equation}
gives all the $H^{int}$'s.
Also, to get rid of the time ordered exponentials we follow the discussions given in pages 338-340 of \cite{Merz}.
This finally gives the expression of the transition amplitude as
\begin{eqnarray}
\begin{array}{rr}
A_{m,n}(t_f,T_2;T_1,t_i) =& e^{i\beta}e^{\frac{i}{\hbar}(E_n-E_m)t_i}\left[D_{m,n}(\xi)
-\frac{i}{\hbar}\theta \{\beta_1 D_{m,n}(\xi)\right. +\beta_2\sqrt{n}D_{m,n-1}(\xi) \\
 &+\beta_2^*\sqrt{n+1}D_{m,n+1}(\xi) 
+\beta_3\sqrt{n(n-1)}D_{m,n-2}(\xi) \\
&\left.+\beta_3^* \sqrt{(n+1)(n+2)}D_{m,n+2}(\xi)\}\right],
\end{array}\label{Amn_final}
\end{eqnarray}
with
\begin{eqnarray}\label{beta_xi}
\xi &=& -\frac{i}{\hbar}\int_{-\infty}^{\infty}d\tau e^{i\omega \tau}z(\tau) \\
\beta &=& \frac{i}{2\hbar^2}\int_{-\infty}^{\infty}d\tau_1\int_{-\infty}^{\infty}d\tau_2\left[ z^*(\tau_1)z(\tau_2)
e^{-i\omega (\tau_1-\tau_2)}-z(\tau_1)z^*(\tau_2)e^{i\omega (\tau_1-\tau_2)}\right] = {\rm real} \\
\beta_1 &=& -m_0\omega \int_{-\infty}^{+\infty}d\tau g^{\prime}(\tau)|\xi(\tau)|^2 +\frac{i}{\hbar}\sqrt{\frac{m_0\hbar\omega}{2}}\int_{-\infty}^{+\infty}d\tau 
\left[ z^{*\prime}(\tau)\xi^2(\tau)e^{-2i\omega \tau}-z^{\prime}(\tau)\xi^{*2}(\tau)e^{2i\omega \tau}\right]  \label{beta1}\\
\beta_2 &=& -m_0\omega \int_{-\infty}^{+\infty}d\tau g^{\prime}(\tau)\xi^*(\tau)
+\frac{2i}{\hbar}\sqrt{\frac{m_0\hbar\omega}{2}}\int_{-\infty}^{+\infty}d\tau z^{*\prime}(\tau)\xi(\tau)e^{-2i\omega \tau}  \\
\beta_3 &=&  \frac{i}{\hbar}\sqrt{\frac{m_0\hbar\omega}{2}}\int_{-\infty}^{+\infty}d\tau z^{*\prime}(\tau)e^{-2i\omega \tau}
\end{eqnarray}
Here the function $\xi(t)$ is given as
\begin{equation} 
\xi(t) = -\frac{i}{\hbar}\int_{-\infty}^{t}d\tau e^{i\omega \tau}z(\tau).
\label{xi_t_0}
\end{equation}
$D_{m,n}(\xi)$'s are the matrix elements of the displacement operator $D(\xi)=e^{-\xi^*a+\xi a^\dagger}$ given by \cite{Perelomov}
\begin{equation}
D_{m,n}(\xi)=\sqrt{\frac{n!}{m!}}e^{-\frac{1}{2}|\xi|^2}\xi^{m-n}L_{n}^{m-n}(|\xi|^2),
\end{equation}
$L_{n}^{k}(x)$ are the associated Laguerre polynomials.
Also, the limits of the integrations have been extended to $-\infty$ and $\infty$ as the integrands are zero in the extended region.
The transition probability is given by
\begin{equation}\label{Pmn}
P_{m,n} =|A_{m,n}(t_f,T_2;T_1,t_i)|^2
\end{equation}
as usual. The arguments have been omitted as the transition probability does not depend on the times $t_f$,$T_2$;$T_1$,$t_i$.

\subsection{$n=0$} \label{n=0}
\begin{figure*}[h]
\begin{center}
\scalebox{1.0}{ \frame{
\begin{tikzpicture}[ultra thick][show background rectangle]
  \draw (-2,0) -- (-1,0) node[below=1pt]{$-T$};
  \draw (-1,0) -- (-1,1);
  \draw (-1,1) -- (1,1) node[right=1pt]{$f_0,g_0$};
  \draw (1,1) -- (1,0);
  \draw (1,0) node[below=1pt]{$+T$} -- (2,0) node[below=1pt]{$t\rightarrow$};
  \begin{scope}[thin]
    \draw[<-] (-3,0) -- (-2,0);
    \draw (-1,0) -- (0,0) node[below=6pt, right=0.2pt]{O};
    \draw (0,0) -- (1,0);
    \draw[->] (2,0) -- (3,0);
    \draw[<-] (0,-1) -- (0,2) node[right=1pt]{$\uparrow f(t),g(t)$};
    \draw (0,2) -- (0,2.5);
    \draw[->] (0,2.5) node[left=50pt]{(1)} -- (0,3);
  \end{scope}
\end{tikzpicture}
}}
\scalebox{1.0}{\includegraphics{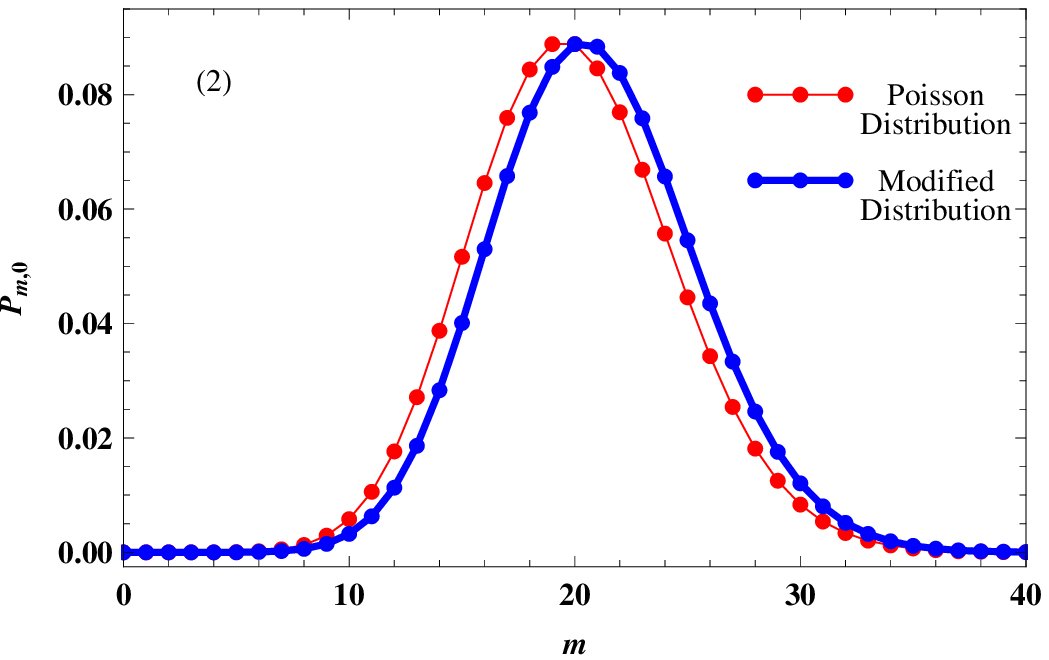}}
\end{center}
\caption{\label{fig_fg} The behaviour of functions $f(t)$ and $g(t)$ with $t$}
\caption{\label{fig_distribution} The modified distribution ($\theta=0.01$) along with the Poisson distribution ($\theta=0$) for
 $m_0=1$, $\omega=1$, $f_0=\sqrt{5}$, $g_0=\sqrt{5}$ and $T=\frac{\pi}{2}$}
\end{figure*}

For the initial state $|\phi_0\rangle$, the transition amplitude is
\begin{equation}
\begin{array}{r}
A_{m,0}(t_f,T_2;T_1,t_i) = e^{i\beta}e^{\frac{i}{\hbar}(E_0-E_m)t_i}e^{-\frac{|\xi|^2}{2}}\frac{\xi^m}{\sqrt{m!}}
\left[ 1-\frac{i}{\hbar}\theta \{ (\beta_1-\beta_2^*\xi^*+\beta_3^*\xi^{* 2}) +m\frac{1}{|\xi|^2}(\beta_2^*\xi^*-2\beta_3^*\xi^{* 2}) \right.\\
\left. +m(m-1)\frac{1}{|\xi|^4}\beta_3^*\xi^{* 2}\}\right],
\label{Am0} 
\end{array}
\end{equation}
and the transition probability becomes (upto linear order in $\theta$)
\begin{equation}
P_{m,0} = |A_{m,0}(t_f,T_2;T_1,t_i)|^2 = e^{-|\xi|^2}\frac{|\xi|^{2m}}{m!}\left[ 1
+\frac{2}{\hbar}\theta\{ A_1+mA_2+m(m-1)A_3\}\right],
\label{Pm0} 
\end{equation}
with
\begin{equation}
A_1 = Im(\beta_2\xi)-Im(\beta_3\xi^2), \quad A_2 = \frac{1}{|\xi|^2}\left(2Im(\beta_3\xi^2)-Im(\beta_2\xi)\right), \quad 
A_3 =-\frac{1}{|\xi|^4}Im(\beta_3\xi^2).
 \label{A3}
\end{equation}
Note that as $m\rightarrow \infty$, the $\theta$-correction starts dominating and in this case the expansion 
upto linear order in $\theta$ is no more meaningful. 
Hence, the above result is valid only for those $m$-values which are far smaller than $1/\sqrt{m_0\omega\theta}$ (in the unit $\hbar=1$). 
For $\theta\rightarrow 0$, the transition probability becomes the well known Poisson distribution as expected. \\
\indent As a specific example let us work with the functions $f(t)$ and $g(t)$ of the form (see Figure \ref{fig_fg})
\begin{equation} \label{fg_expression}
\left.
\begin{array}{rcl}
 f(t) &=& f_0\left[\Theta (t+T)-\Theta (t-T)\right] \\
 g(t) &=& g_0\left[\Theta (t+T)-\Theta (t-T)\right]
\end{array}
\right\rbrace \quad\quad ;T>0.
\end{equation}
For these functions we get 
\begin{eqnarray}
 A_1 &=& \frac{2f_0}{m_0\omega^3}(f_0^2+m_0^2\omega^2 g_0^2)\sin^2{\omega T}\cos{2\omega T} \\
A_2 &=& m_0\omega g_0 \sin{2\omega T}-f_0\cos{2\omega T} \\
A_3 &=& -\frac{m_0^2\omega^4 g_0 \cot{\omega T}}{f_0^2+m_0^2\omega^2g_0^2}.
\end{eqnarray}
Now the choices $m_0=1$, $\omega=1$, $f_0=\sqrt{5}$, $g_0=\sqrt{5}$ and $T=\frac{\pi}{2}$ in commutative case ($\theta=0$) give the 
following Poisson distribution: $P_{m,0}=e^{-20}\frac{20^m}{m!}$,
while for nonzero $\theta$ the probability distribution modifies to $P_{m,0}=e^{-20}\frac{20^m}{m!}[1+2\theta\sqrt{5}(m-20)]$.
The $\theta$-correction becomes of the order of the $\theta$-independent part when $m$ approaches the value 
$\tilde{m}(\theta)=\left(20+\frac{1}{2\sqrt{5}\theta}\right)$. 
Hence, our result is valid only in the region where $m<\tilde{m}(\theta)$.
Note that $\tilde{m}(\theta)\sim\frac{1}{\theta}$ rather than $\frac{1}{\sqrt{\theta}}$ 
because the $A_3$ is identically zero for the choices taken.
We choose $\theta=0.01$ ($\tilde{m}(\theta=0.01)\approx 42$) and get
 \begin{equation}
 P_{m,0}=e^{-20}\frac{20^m}{m!}[1+0.02\sqrt{5}(m-20)].
\end{equation}
This deformed distribution along with the Poisson distribution is shown in Figure \ref{fig_distribution}.
Such deformation of the Poisson distribution suggests that the vacuum does not evolve to be a coherent state anymore.
To explore this further let us look at the time-evolution of position and momentum uncertainties.

\section{The time evolution of $\Delta_x$ and $\Delta_p$}\label{uncertainties}
\begin{figure*}[h]
\begin{center}
\scalebox{0.7}{\includegraphics{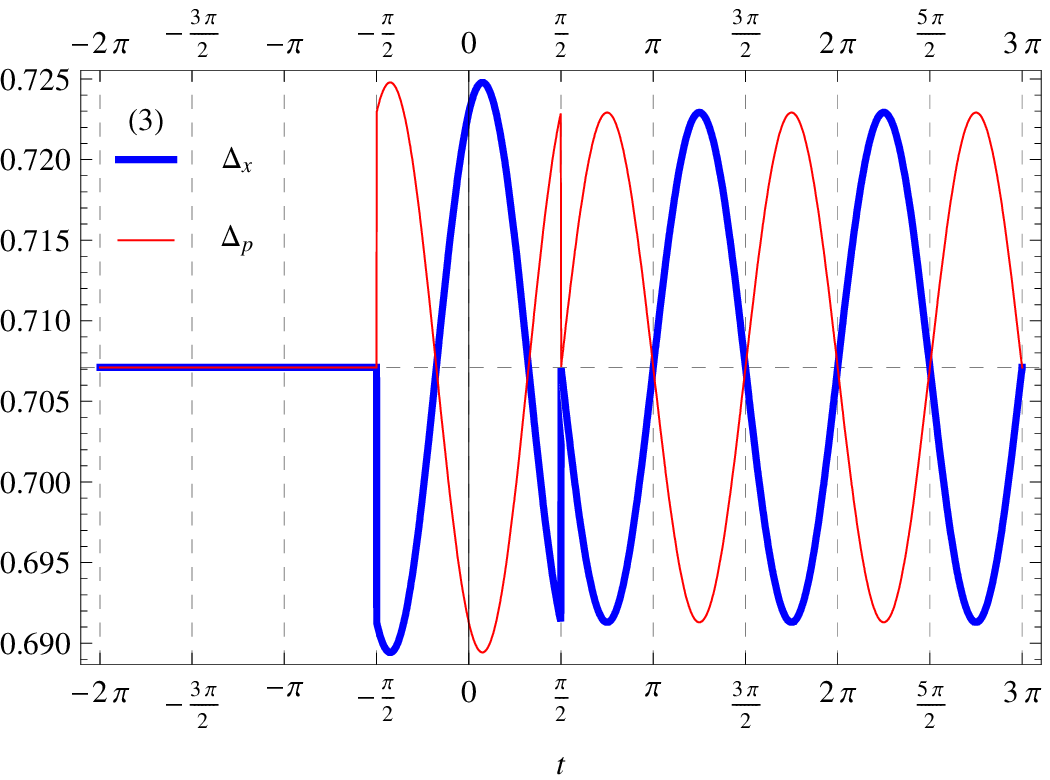}}
\scalebox{0.7}{\includegraphics{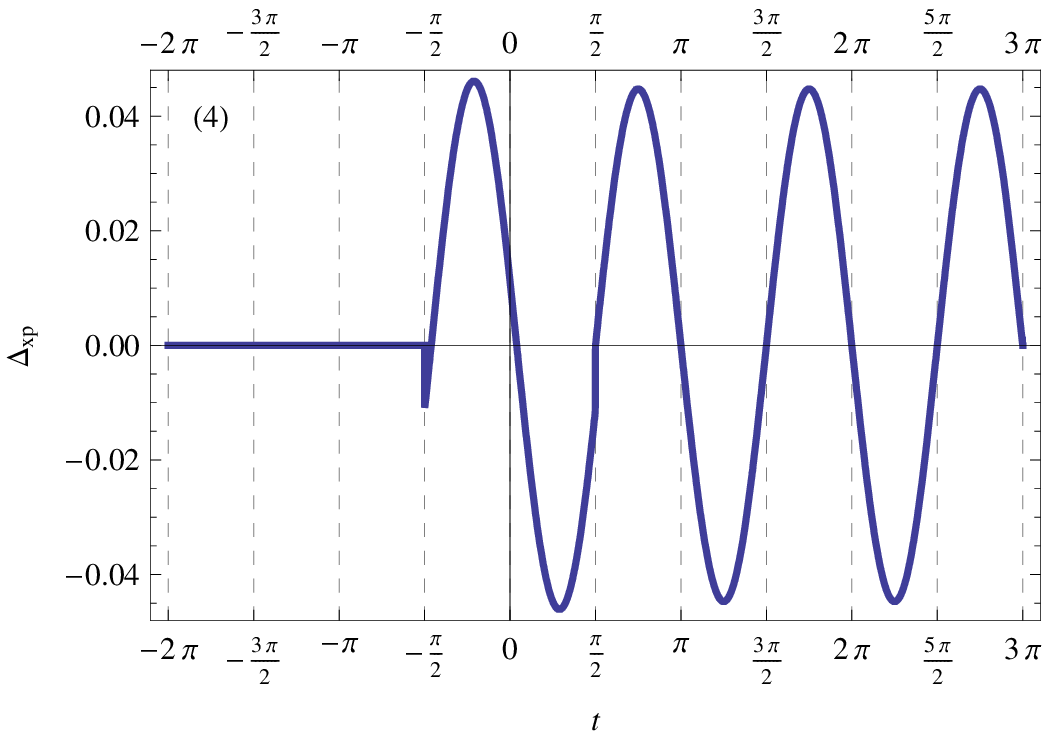}}
\end{center}
\caption{\label{fig_deltaxdeltap} The time-dependences of the uncertainties $\Delta_{x}$ and $\Delta_{p}$ 
for $m_0=1$, $\omega=1$, $f_0=\sqrt{5}$, $g_0=\sqrt{5}$, $T=\frac{\pi}{2}$ and $\theta=0.01$}
\caption{\label{fig_deltaxp} The time-dependence of $\Delta_{xp}$ for the same choice of values}
\end{figure*}

The expectation value of any operator $\hat{\mathcal{O}}$ in a state $\hat{\psi}(\hat{x}_0,\hat{x}_1)$ at any time $t$ is defined to be
\begin{equation}
 \langle \hat{\mathcal{O}} \rangle_t =  \left(\hat{\psi},\hat{\mathcal{O}}\hat{\psi}\right)_t.
\end{equation}
Also,
\begin{equation*}
 \langle \hat{\mathcal{O}} \rangle_{t+\tau} =  \left(\hat{\psi},\hat{\mathcal{O}}\hat{\psi}\right)_{t+\tau} =
\left(\hat{\psi}(\hat{x}_0+\tau,\hat{x}_1),\hat{\mathcal{O}}\hat{\psi}(\hat{x}_0+\tau,\hat{x}_1)\right)_t.
\end{equation*}
Thus the time evolution of the expectation value of an operator is given by that of the state in which it is being calculated.
For the QFHO in $\mathds{R}_\theta^{1,1}$ the time evolution of any operator $\hat{\mathcal{O}}$ will be given by
\begin{equation}
\frac{d}{dt}\langle \hat{\mathcal{O}}\rangle=\frac{\partial}{\partial t}\langle \hat{\mathcal{O}}\rangle
+\frac{i}{\hbar}\left\langle\left[H(t),\hat{\mathcal{O}}\right]\right\rangle,
\end{equation} 
where $H(t)$ is the Hamiltonian (\ref{h}).
The uncertainty in any observable $\hat{\mathcal{O}}$ is given by
\begin{equation}
\Delta_{\mathcal{O}}=\sqrt{\langle\hat{\mathcal{O}}^2\rangle-\langle\hat{\mathcal{O}}\rangle^2}.
\end{equation} 
Thus the evolution of $\Delta_x^2$ and $\Delta_p^2$ is
\begin{equation}\left.
\begin{array}{rcl}
\frac{d}{dt}\Delta_x^2 &=& 2\left(\frac{1}{m_0}-\frac{\theta}{\hbar} g^{\prime}(t)\right)\left(\frac{1}{2}\langle xp+px\rangle-\langle x\rangle\langle p\rangle\right)-\frac{2\theta}{\hbar} f^{\prime}(t)\Delta_x^2 \\
\frac{d}{dt}\Delta_p^2 &=& -2m_0\omega^2\left(\frac{1}{2}\langle xp+px\rangle-\langle x\rangle\langle p\rangle\right)+\frac{2\theta}{\hbar} f^{\prime}(t)\Delta_p^2 
\end{array}\right\rbrace.
\end{equation}
Defining
\begin{equation}
\Delta_{xp}=\frac{1}{2}\langle xp+px\rangle-\langle x\rangle\langle p\rangle,
\end{equation}
we find the following first order coupled equations
\begin{equation}\left.
\begin{array}{rcl}
\frac{d}{dt}\Delta_x^2 &=& 2\left(\frac{1}{m_0}-\frac{\theta}{\hbar} g^{\prime}(t)\right)\Delta_{xp}-\frac{2\theta}{\hbar} f^{\prime}(t)\Delta_x^2 \\
\frac{d}{dt}\Delta_p^2 &=& -2m_0\omega^2\Delta_{xp}+\frac{2\theta}{\hbar} f^{\prime}(t)\Delta_p^2 \\
\frac{d}{dt}\Delta_{xp} &=& \left(\frac{1}{m_0}-\frac{\theta}{\hbar} g^{\prime}(t)\right)\Delta_p^2 -m_0\omega^2\Delta_x^2 
\end{array}\right\rbrace.
\end{equation}
As the initial state is the vacuum, the initial conditions for the above are
\begin{equation}
\Delta_x^2(t\rightarrow -\infty) = \frac{\hbar}{2m_0\omega}, \quad
\Delta_p^2(t\rightarrow -\infty) = \frac{m_0\hbar\omega}{2}, \quad
\Delta_{xp}(t\rightarrow-\infty) = 0 .
\end{equation} 
Our strategy for solving these equations is simple.
We do so perturbatively in $\theta$.
A straightforward computation gives
\begin{equation}\left.
\begin{array}{c}
\Delta_x(t) = \sqrt{\frac{\hbar}{2m_0\omega}}-\frac{\theta}{2}\sqrt{\frac{m_0}{2\hbar\omega}}
\left[\frac{2}{m_0}f(t)+\frac{4\omega}{m_0}\displaystyle{\int_{-\infty}^t}d\tau \sin\{2\omega(\tau-t)\}f(\tau)
+2\omega^2 \displaystyle{\int_{-\infty}^t}d\tau \cos\{2\omega(\tau-t)\}g(\tau)\right] \\
\\
\Delta_p(t) = \sqrt{\frac{m_0\hbar\omega}{2}}+\frac{\theta m_0}{2}\sqrt{\frac{m_0\omega}{2\hbar}}
 \left[\frac{2}{m_0}f(t)+\frac{4\omega}{m_0}\displaystyle{\int_{-\infty}^t}d\tau \sin\{2\omega(\tau-t)\}f(\tau)
+2\omega^2 \displaystyle{\int_{-\infty}^t}d\tau \cos\{2\omega(\tau-t)\}g(\tau)\right]\\
\\
\Delta_{xp}(t) = \theta\left[-\frac{m_0\omega}{2}g(t)+2\omega\displaystyle{\int_{-\infty}^t}d\tau \cos\{2\omega(\tau-t)\}f(\tau)
-m_0\omega^2 \displaystyle{\int_{-\infty}^t}d\tau \sin\{2\omega(\tau-t)\}g(\tau)\right]
\end{array}\right\rbrace.
\label{uncertainties_final2}
\end{equation}
The fundamental uncertainty product (to linear order in $\theta$) is
\begin{equation}
\Delta_x(t).\Delta_p(t)=\frac{\hbar}{2}.
\end{equation}
Thus the vacuum state evolves to a ``squeezed state'' rather than a coherent state as in the commutative case \cite{Walls}.  
The uncertainties in the commutative case depend only on the product $m_0\omega$.
But, their $\theta$-corrections change with $\omega$ even if $m_0\omega$ is kept constant.
Also, the squeezing effect is oscillatory in time as is obvious from the $\theta$-dependent terms in (\ref{uncertainties_final2}).
For the specific forms of $f(t)$ and $g(t)$ of (\ref{fg_expression}) we get
\begin{equation}
\Delta_{x}(t)=\left\{
 \begin{array}{l}
  \sqrt{\frac{\hbar}{2m_0\omega}} \hspace*{11 cm} ;t<-T \\
\sqrt{\frac{\hbar}{2m_0\omega}}-\frac{\theta}{2}\sqrt{\frac{1}{2\hbar m_0\omega}} 
\left[2f_0\cos\{2\omega(t+T)\}+m_0g_0\omega \sin\{2\omega(t+T)\}\right] \quad  ;-T<t<T \\
\begin{array}{l}
\sqrt{\frac{\hbar}{2m_0\omega}}-\frac{\theta}{2}\sqrt{\frac{1}{2\hbar m_0\omega}} 
\left[2f_0\left(\cos\{2\omega(t+T)\}-\cos\{2\omega(t-T)\}\right)\right. \\
\hspace*{3 cm} +\left. m_0g_0\omega \left(\sin\{2\omega(t+T)\}-\sin\{2\omega(t-T)\}\right)\right] \quad \quad  
 \end{array} ;t>T,
 \end{array}\right.
\end{equation}
\begin{equation}
\Delta_{p}(t)=\left\{
 \begin{array}{l}
  \sqrt{\frac{m_0\hbar\omega}{2}} \hspace*{11 cm} ;t<-T \\
\sqrt{\frac{m_0\hbar\omega}{2}}+\frac{\theta}{2}\sqrt{\frac{m_0\omega}{2\hbar}} 
\left[2f_0\cos\{2\omega(t+T)\}+m_0g_0\omega \sin\{2\omega(t+T)\}\right] \quad  ;-T<t<T \\
\begin{array}{l}
\sqrt{\frac{m_0\hbar\omega}{2}}+\frac{\theta}{2}\sqrt{\frac{m_0\omega}{2\hbar}} 
\left[2f_0\left(\cos\{2\omega(t+T)\}-\cos\{2\omega(t-T)\}\right)\right. \\
\hspace*{3 cm} +\left. m_0g_0\omega \left(\sin\{2\omega(t+T)\}-\sin\{2\omega(t-T)\}\right)\right] \quad \quad  
 \end{array} ;t>T
 \end{array}\right.
\end{equation}
\begin{equation}
{\rm and} \quad \quad \Delta_{xp}(t)=\left\{
 \begin{array}{l}
  0 \hspace*{9 cm} ;t<-T \\
\frac{\theta}{2}\left[2f_0\sin\{2\omega(t+T)\}-m_0\omega g_0\cos\{2\omega(t+T)\}\right] \quad  ;-T<t<T \\
\begin{array}{l}
\frac{\theta}{2} \left[2f_0\left(\sin\{2\omega(t+T)\}-\sin\{2\omega(t-T)\}\right)\right. \\
\left. \quad -m_0\omega g_0\left(\cos\{2\omega(t+T)\}-\cos\{2\omega(t-T)\}\right)\right]
\end{array} \quad ;t>T.
 \end{array}\right.
\end{equation}
For the same choice of parameters as in the last section we get
\begin{equation}
\Delta_{x}(t)=\left\{
\begin{array}{l}
 \frac{1}{\sqrt{2}} \hspace*{5 cm} ;t<-\frac{\pi}{2} \\
\frac{1}{\sqrt{2}} + 0.01 \sqrt{\frac{5}{2}}\left(\cos{2t}+\frac{1}{2}\sin{2t}\right) \quad ;-\frac{\pi}{2}<t<\frac{\pi}{2} \\
\frac{1}{\sqrt{2}} + 0.01 \sqrt{\frac{5}{2}}\sin{2t} \quad ;t>\frac{\pi}{2},
\end{array}
\right.
\end{equation}

\begin{equation}
\Delta_{p}(t)=\left\{
\begin{array}{l}
 \frac{1}{\sqrt{2}} \hspace*{5 cm} ;t<-\frac{\pi}{2} \\
\frac{1}{\sqrt{2}} - 0.01 \sqrt{\frac{5}{2}}\left(\cos{2t}+\frac{1}{2}\sin{2t}\right) \quad ;-\frac{\pi}{2}<t<\frac{\pi}{2} \\
\frac{1}{\sqrt{2}} - 0.01 \sqrt{\frac{5}{2}}\sin{2t} \quad ;t>\frac{\pi}{2}
\end{array}
\right.
\end{equation}

\begin{equation}
{\rm and} \quad \quad \Delta_{xp}(t)=\left\{
\begin{array}{l}
 0 \hspace*{5 cm} ;t<-\frac{\pi}{2} \\
0.01 \sqrt{5}\left(\frac{1}{2}\cos{2t}-\sin{2t}\right) \quad ;-\frac{\pi}{2}<t<\frac{\pi}{2} \\
-0.02 \sqrt{5}\sin{2t} \quad ;t>\frac{\pi}{2}.
\end{array}
\right.
\end{equation}
Figures \ref{fig_deltaxdeltap} and \ref{fig_deltaxp} show the time-dependence of the different uncertainties.
The discontinuities at $t=\pm \frac{\pi}{2}$ is simply the manifestation of the fact that the functions $f(t)$ and $g(t)$ 
themselves are discontinuous at these times.
Before the interaction was switched on, the uncertainties were having values equal to those for the vacuum state.
During the time of nonvanishing interaction (and even after the interaction gets switched off!), 
they oscillate with frequency equal to twice that of the oscillator.

\section{Implications in Quantum Optics}\label{quantum_optics}
In Quantum Optics a monochromatic (single-mode) coherent light field is usually described by the harmonic oscillator coherent states \cite{Glauber:1963tx}.
It has also been shown that a coherent state (in particular the vacuum state)
remains to be coherent under the FHO Hamiltonian \cite{Carruthers}.
The annihilation and creation operators for photons are related to the field quadratures $X_1$ and $X_2$ by
\begin{equation}
a = X_1+iX_2, \quad  a^{\dagger} = X_1-iX_2,
\end{equation}
$X_1$ and $X_2$ being hermitian. 
The commutation $\left[a,a^{\dagger}\right]=1$ translates to $\left[X_1,X_2\right]=\frac{i}{2}$.
The coherent state has different uncertainties as $\Delta_{X_1} = \frac{1}{2}$, $\Delta_{X_2} = \frac{1}{2}$ and
 $\Delta_{X_1X_2} = 0$ $\Rightarrow$ $\Delta_{X_1}.\Delta_{X_2}=\frac{1}{4}$ which is the minimum. 
Also, the photon count (probability for having a certain number of photons) in the coherent state is given by 
the transition probabilities of the corresponding number eigenstate and the profile is Poissonian. \\
\indent The FHO Hamiltonian
\begin{eqnarray}
H(t) &=& \hbar\omega(X_1^2+X_2^2)+\sqrt{\frac{2\hbar}{m_0\omega}}f(t)X_1+\sqrt{2\hbar m_0\omega}g(t)X_2 \nonumber \\
&=& \hbar\omega\left(a^{\dagger}a+\frac{1}{2}\right)+z^*(t)a+z(t)a^{\dagger}
\end{eqnarray}
($z(t)$ is related with $f(t)$ and $g(t)$ by (\ref{z})) with the effective noncommutativity 
between time and the field quadrature $X_1$ of the form
\begin{equation}
\left[t,X_1\right]=i\sqrt{\frac{m_0\omega}{2\hbar}}\theta \label{QO_com}
\end{equation}
will allow us to use the calculation of the previous sections.
The photon count will be given by (\ref{Pm0}), while the uncertainties in the field quadratures will get modified as
\begin{equation}\left.
\begin{array}{rcl}
\Delta_{X_1}(t) &=& \frac{1}{2}+\frac{\theta m_0}{4\hbar}\displaystyle{\int_{-\infty}^t}d\tau \sin\left(2\omega(\tau-t)\right)\left[\omega g^{\prime}(\tau)-\frac{1}{m_0\omega}f^{\prime\prime}(\tau)\right] \\
\Delta_{X_2}(t) &=& \frac{1}{2}-\frac{\theta m_0}{4\hbar}\displaystyle{\int_{-\infty}^t}d\tau \sin\left(2\omega(\tau-t)\right)\left[\omega g^{\prime}(\tau)-\frac{1}{m_0\omega}f^{\prime\prime}(\tau)\right] \\
\Delta_{X_1X_2}(t) &=& \frac{\theta}{2\hbar}\left(\frac{1}{2\omega}f^{\prime}(t)-\frac{m_0}{2}\displaystyle{\int_{-\infty}^t}d\tau \cos\{2\omega(\tau-t)\}\left[\omega g^{\prime}(\tau)+\frac{1}{m_0\omega}f^{\prime\prime}(\tau)\right]\right)
\end{array}\right\rbrace.
\end{equation}
We further study the correlation among the photons. The time-evolved vacuum state
\begin{equation}
|i(t\rightarrow\infty)\rangle=\displaystyle{\sum_{m=0}^{\infty}}A_{m,0}|m\rangle
\end{equation}
will give
\begin{equation}
\bar{N}=\langle i(t\rightarrow\infty)|a^{\dagger}a|i(t\rightarrow\infty)\rangle=\displaystyle{\sum_{m=1}^{\infty}}mP_{m,0}
= |\xi|^2-\frac{2\theta}{\hbar^2}Im(\beta_2\xi),
\end{equation}
$\bar{N}$ being the average number of photons in state $|i(t\rightarrow\infty)\rangle$. Also
\begin{equation}
\langle i(t\rightarrow\infty)|a^{\dagger}a^{\dagger}aa|i(t\rightarrow\infty)\rangle=\displaystyle{\sum_{m=2}^{\infty}}m(m-1)P_{m,0}
= |\xi|^4-\frac{4\theta}{\hbar^2}\left(Im(\beta_3\xi^2)+|\xi|^2 Im(\beta_2\xi)\right).
\end{equation}
This, to linear order in $\theta$, gives the 2nd order correlation among photons with zero time delay to be equal to 
(see Appendix \ref{correlation})
\begin{equation}
g^{(2)}(0) = 1-\frac{\frac{4\theta}{\hbar^2} Im(\beta_3\xi^2)}{\left(|\xi|^4-\frac{4\theta}{\hbar^2}|\xi|^2 Im(\beta_2\xi)\right)}
= 1-\frac{\frac{4\theta}{\hbar^2} Im(\beta_3\xi^2)}{\bar{N}^2}.
\end{equation}
For the case $Im(\beta_3\xi^2)<0\Rightarrow g^{(2)}(0)>1$, the photons try to bunch together 
while for $Im(\beta_3\xi^2)>0\Rightarrow g^{(2)}(0)<1$, they anti-bunch \cite{Walls_book}.
For the functions (\ref{fg_expression}), we get
\begin{equation}
 Im(\beta_3\xi^2) = \frac{2(f_0^2+m_0^2\omega^2g_0^2)\sin^2 \omega T}{\omega^2}g_0\sin 2\omega T,
\end{equation}
which implies that the bunching or anti-bunching will depend only on the sign of the factor $g_0\sin 2\omega T$.
For the choices taken in figures \ref{fig_distribution}, \ref{fig_deltaxdeltap} $\&$ \ref{fig_deltaxp}, $\omega T = \frac{\pi}{2}$
and hence no bunching or anti-bunching occurs.

\section{Conclusions} \label{conclusions}
In this paper we developed a formalism to compute the transitions between states 
of a quantum mechanical system with noncommutative time.
We found that for a free Hamiltonian in $\mathds{R}_\theta^{1,1}$ which is independent of time, the transitions are equal to
the same for a different Hamiltonian in $\mathds{R}^{1,1}$ found after the replacements (\ref{replace}).
The time evolution of an operator and its expectation value (and hence also its uncertainty) can also be found in a similar manner.
Specifically, for FHO the transition probabilities get modified and is given by (\ref{Amn_final}) and (\ref{Pmn}).
The Poissonian distribution for the ``vacuum to any state transition'' also gets modified and is given by (\ref{Pm0}).
The study of uncertainties in position and momentum says that the time-evolved state is no more coherent.
It gets some squeezing effect due to the noncommutativity, keeping the product of the uncertainties minimum.
These uncertainties are explicitly found and is given in (\ref{uncertainties_final2}).
The leading order corrections in these uncertainties are oscillatory in time and they depend independently on the mass of the
particle $m_0$ and the frequency of the oscillator $\omega$ (note that the commutative uncertainties depend only on the product $m_0\omega$). 
These results suggest a possible modeling of the noncommutativity for the nonlinear phenomena in Quantum Optics.
The noncommutativity results in the following nonlinear effects:
\begin{enumerate}
\item{The photon-count gets modified from the usual Poisson distribution.}
\item{The uncertainties in the field quadratures change keeping the product minimum (the squeezing effect).}
\item{The second order correlation function $g^{(2)}(0)$ gets modified producing new effects 
like bunching or anti-bunching of photons depending on the value of $Im(\beta_3\xi^2)$.}
\end{enumerate}
All these observations suggest that the noncommutativity produces incoherency in the otherwise coherent field.

As a future work one can try to formulate the scattering process in higher dimensions and study its implications in quantum optics. The correspondence found in this paper between noncommutativity and quantum optics also encourages one to study such possibilities in other forms of time-sapce noncommutativity. As an example one can start with assuming the spacetime dependent noncommutative parameter $\theta$ \cite{Majid:1994cy, Sitarz:1994rh, Lukierski:1993wx}.

\section*{Appendix}
\appendix
\section{The Green's operator function} \label{greens}
Expanding $G(t,t_0)$ as a Fourier integral we get
\begin{equation*}
G(t,t_0) = G(t-t_0)
= \frac{1}{2\pi} \int_{-\infty}^{+\infty}d\omega^{\prime}g(\omega^{\prime})e^{-i\omega^{\prime}(t-t_0)},
\end{equation*}
where $g(\omega^{\prime})$ is given by $g(\omega^{\prime})=(\hbar\omega^{\prime}-H_0)^{-1}$.
Introducing the simple harmonic eigenstate basis and using the completeness relation we get
\begin{equation*}
G(t,t_0)=\sum_{j=0}^{\infty}\left(\lim_{\epsilon\rightarrow 0^+} \frac{1}{2\pi}\int_{-\infty}^{+\infty}
e^{-i\omega^{\prime}(t-t_0)}\frac{1}{\hbar(\omega^{\prime}-\frac{E_n}{\hbar}+i\epsilon)}\right) |\phi_j\rangle\langle\phi_j|.
\end{equation*}
Here, $i\epsilon$ has been introduced to avoid the pole on the contour (real axis).
After finding the integral inside the summation using the complex analysis we get the Green's function to be
\begin{equation}
G(t,t_0) = -\frac{i}{\hbar}\Theta (t-t_0) e^{-i\frac{H_0}{\hbar}(t-t_0)} \label{G_sol},
\end{equation}
where $\Theta(x)$ is the Heaviside Step Function.
\section{Simplifying $A^{(1)}(T_2,T_1)$} \label{simplify_A1}
To simplify (\ref{A1_series}) what we do is to find the differential equation for $A^{(1)}(t,T_1)$ with independent variable $t$ and solve it with proper initial conditions.
The differential equation has been found to be
\begin{equation}
\left( i\hbar\frac{\partial}{\partial t}-H_{I0}^{int}(t)\right)A^{(1)}(t,T_1)=H_{I1}^{int}(t)A^{(0)}(t,T_1),
\end{equation}
which is a first order equation and hence $A^{(1)}(t,T_1)$ is unique if an initial condition is given.
The initial condition comes from the fact that  $A^{(0)}(t,T_1)$ and $A^{(1)}(t,T_1)$ must become identity and zero respectively for $t=T_1$, i.e., no interaction.
Thus, $A^{(1)}(T_1,T_1)=0$.
To solve the equation we define the Green's operator function $G_{int}(t,t_0)$ as
\begin{equation}
\left( i\hbar\frac{\partial}{\partial t}-H_{I0}^{int}(t)\right)G_{int}(t,t_0)=\delta(t-t_0).
\label{Gint}
\end{equation}
Now generalizing solution (\ref{G_sol}) for the time-dependent case we get
\begin{equation}
G_{int}(t,t_0) = -\frac{i}{\hbar}\Theta (t-t_0)T\left[ e^{-\frac{i}{\hbar}\int_{t_0}^{t}d\tau H_{I0}^{int}(\tau)}\right]
= -\frac{i}{\hbar}\Theta (t-t_0)A^{(0)}(t,t_0).
\label{Gint_sol}
\end{equation}
It can be easily checked that the above expression for the $G_{int}(t,t_0)$ satisfies the corresponding differential equation. 
The solution for $A^{(1)}(t,T_1)$ is then given by
\begin{equation}
A^{(1)}(t,T_1)=A_{hom}^{(1)}(t,T_1)+\int_{-\infty}^{+\infty}dt_0\,\,G_{int}(t,t_0)H_{I1}^{int}(t_0)A^{(0)}(t_0,T_1),
\end{equation}
where $A_{hom}^{(1)}(t,T_1)$ is the solution of the homogeneous equation
$\left( i\hbar\frac{\partial}{\partial t}-H_{I0}^{int}(t)\right)A_{hom}^{(1)}(t,T_1)=0$.
The $\Theta$-function in the expression of $G_{int}(t,t_0)$
and the fact that interaction was off before $T_1$, with the initial condition for $A^{(1)}(T_1,T_1)=0$,
gives the initial condition for $A_{hom}^{(1)}(t,T_1)$, i.e., $A_{hom}^{(1)}(T_1,T_1)=0$.
The only solution of the homogeneous equation with this initial condition is $A_{hom}^{(1)}(t,T_1)=0$.
Thus we get
\begin{equation}
A^{(1)}(t,T_1)=\int_{-\infty}^{+\infty}dt_0 \,\,G_{int}(t,t_0) H_{I1}^{int}(t_0)A^{(0)}(t_0,T_1).
\end{equation}
Now, putting the expression of $G_{int}(t,t_0)$ above and 
introducing $A^{(0)}(t_0,T_1)\left[A^{(0)}(t_0,T_1)\right]^{-1}$ before $H_{I1}^{int}(t_0)$, we get (\ref{A1_A0}).
Here we have also used the following property of $A^{(0)}(t,T_1)$:
\begin{equation*}
 A^{(0)}(t,t_0)A^{(0)}(t_0,T_1)=A^{(0)}(t,T_1).
\end{equation*}

\section{The correlation function} \label{correlation}
An operator corresponding to the detection of a photon by a detector 
should be proportional to the annihilation operator $a$  (say $ka$) \cite{Walls_book, Glauber:1963fi}.
Hence if $|i\rangle$ is the initial state of the radiation field, the state after the detection of one photon is $ka|i\rangle$.
The amplitude for going to the final state $|f\rangle$ is given by $k\langle f|a|i\rangle$.
The corresponding probability is $|k|^2|\langle f|a|i\rangle|^2$.
Thus the probability of detection of one photon in the state $|i\rangle$
\begin{equation}
P_1 = \displaystyle{\sum_f}|k|^2|\langle f|a|i\rangle|^2
= |k|^2\displaystyle{\sum_f}\langle i|a^{\dagger}|f\rangle\langle f|a|i\rangle
= |k|^2\langle i|a^{\dagger}a|i\rangle.
\end{equation}
Similarly, probability of detection of two photons with a time delay of $\tau$ is
\begin{equation}
\begin{array}{r}
P_2 = \displaystyle{\sum_f}|k|^4|\langle f|a(t+\tau)a(t)|i\rangle|^2
= |k|^4\displaystyle{\sum_f}\langle i|a^{\dagger}(t)a^{\dagger}(t+\tau)|f\rangle\langle f|a(t+\tau)a(t)|i\rangle \\
= |k|^4\langle i|a^{\dagger}(t)a^{\dagger}(t+\tau)a(t+\tau)a(t)|i\rangle.
\end{array}
\end{equation}
The 2nd order correlation function with a time delay $\tau$ is defined as 
\begin{equation}
g^{(2)}(\tau) = \frac{P_2(\tau)}{P_1^2}
= \frac{\langle i|a^{\dagger}(t)a^{\dagger}(t+\tau)a(t+\tau)a(t)|i\rangle}{\langle i|a^{\dagger}(t)a(t)|i\rangle^2}.
\end{equation}
For $\tau=0$
\begin{equation}
g^{(2)}(0)= \frac{\langle i|a^{\dagger}(t)a^{\dagger}(t)a(t)a(t)|i\rangle}{\langle i|a^{\dagger}(t)a(t)|i\rangle^2} 
=  \frac{\langle i(t)|a^{\dagger}a^{\dagger}aa|i(t)\rangle}{\langle i(t)|a^{\dagger}a|i(t)\rangle^2},
\label{g20}
\end{equation}
For a coherent state it can be calculated to be equal to 1.

\section*{Acknowledgement} It is a pleasure to thank Sachindeo Vaidya for introducing me to this problem,
 for all the useful discussions while solving it and for useful comments on the manuscript. 
I also want to thank Nirmalendu Acharyya and Sandeep Chatterjee for
going through the manuscript and suggesting some welcome changes.

\end{document}